\documentclass[11pt]{article}
\usepackage{amsmath}
\usepackage{amssymb}
\usepackage{authblk}
\usepackage[super,compress]{cite}

\usepackage{graphicx}
\usepackage{caption}

\def\be{\begin{equation}}
\def\ee{\end{equation}}
\def\ba{\begin{array}}
\def\bacc{\begin{array} {cc}}
\def\ea{\end{array}}
\def\bea{\begin{eqnarray}}
\def\eea{\end{eqnarray}}
\def\bd{\begin{displaymath}}
\def\ed{\end{displaymath}}

\def\nn{\nonumber}

\newcommand{\te}{\theta^+}
\newcommand{\bt}{\bar{\theta}^+}
\newcommand{\I}{{\int\text{d}\te\text{d}\bar{\theta}^+}}
\newcommand{\D}{\mathcal{D}}

\newcommand{\F}{\widetilde{\mathcal{F}}}

\newcommand{\T}{\mathcal{T}}
\newcommand{\Tr}{\text{Tr}}
\newcommand{\BLGamma}{\mathbf{\Gamma}}
\newcommand{\BLPhi}{\mathbf{\Phi}} 
\newcommand{\BLE}{\mathbf{\widetilde E}}

\newcommand{\BLF}{\widetilde{\mathbf{\mathcal{F}}}}

\DeclareMathOperator{\tr}{tr}

\begin{document}
\author{Nana Cabo Bizet}

\affil{ \it Department of Physics, University of Guanajuato,\\ \it  Loma del Bosque 103, León, Guanajuato, México}

%
%

\title{T-dualities in gauged linear sigma models}




\maketitle

\begin{abstract}

We describe non-Abelian T-dualities for $\mathcal{N} = 2$ two dimensional gauged linear sigma model (GLSM). We start with the case of left and right $(2, 2)$ supersymmetry (SUSY),  $U(1)$ gauge group, and global non-Abelian symmetries. Our analysis applies to the specialization of the GLSM with the global group $SU(2)\times SU(2)$, whose original model is the resolved conifold.  We analyze the dual model and compute the periods on the dual geometry, matching the effective potential for the $U(1)$ gauge field, and discuss the coincident singularity at the conifold point. We further present  the non-Abelian T-dualization of models with $(0,2)$ SUSY, analyzing the case of global symmetry $SU(2)$. In this case the full non-Abelian duality can be solved. The dual geometries with and without
non-perturbative corrections to the superpotential coincide. The structure of the dual non-perturbative corrections employed are determined based on symmetry arguments.  The non-Abelian T-dualities reviewed here lead to potential new symmetries between physical theories.


\end{abstract}


\section{Introduction}

String theory possess symmetries and dualities, which relate several of its corners. In particular, T-duality
constitutes a discovery, first developed by Buscher \cite{buscher1, buscher2}, that relates a theory on a circle with radius $R$, with a theory on a circle of radius $\alpha'/R$. The spectrum of physical states is matched between T-dual  string theories, meaning that winding modes and momentum modes are exchanged. There is a non-Abelian version of T-duality, when the compact dimensions posses global non-Abelian symmetries. In the non linear sigma model (NLSM) i.e. string theory, the equations of motion are exchanged with the Bianchi identities, leading to equivalent theories \cite{ossaquevedo,giveonrocek}. On the context of NLSM non-Abelian T-duality is a more complex construction, and there is still discussion on to what extend models obtained out of it are equivalent \cite{giveonrocek}. In particular, is not possible in general to find dual models that posses the global non-Abelian symmetries of the original model, and therefore is not clear how to obtain the original model back. Some of these problems have been adressed in the context of Poisson-Lie T-duality \cite{poisson1,poisson2,poisson3}.

On the other hand Mirror Symmetry (MS) enunciates that for each Calabi-Yau (CY) variety, there is a dual geometry on which the  complex structure moduli space and the K\"ahler moduli space are exchanged \cite{candelasossa}.  MS  maps type IIB string theory on a given CY manifold with type IIA string theory
on the mirror CY. As well the relation between MS and T-duality has been explored, in the context 
of string theory \cite{syz} and also by looking at the UV description of strings, given by the GLSM \cite{morrisonplesser,horivafa}. It has been established that Abelian T-duality in GLSMs  leads to MS between target spaces in the seminal works of Morrison and Plesser \cite{morrisonplesser} and Hori and Vafa \cite{horivafa}. Employing the duality procedure of Roceck and Verlinde \cite{rocekverlinde} we have extended the mentioned exploration\cite{horivafa} by constructing general T-dual models in GLSMs and in particular Non-Abelian T-dual models \cite{cmps18,cs20,cjs22,cdg23}. One motivation is to explore the wide group of T-dualities, and to develop tools that serve to describe MS and further symmetries for non Complete Intersection Calabi-Yau (nCICY)\cite{horitong,determinantal,gusharpe5}. The starting point of this exploration are Abelian GLSMs, which possess non-Abelian global symmetries. In this article we review the construction of non-Abelian T-duality originally performed in
models with two left SUSYs ($N_L$) and two right SUSYs ($N_R$) i.e.  $(N_L,N_R)=(2,2)$ \cite{cmps18} , and recently for $(N_L,N_R)=(2,0)$ SUSY \cite{cdg23}. We explain the equations of motion that lead to the dual theories, on the examples analyzed so far. As a new development, we describe how one can compute the periods in te dual geometries, and we do so for the non-Abelian dual of the resolved conifold. We also discuss our plan to test further these new  potential string symmetries.

In Section \ref{sec2} we present the basics of supersymmetric $(2,2)$ GLSMs, Section \ref{sec3} is devoted to implement the Abelian T-duality procedure in $(2,2)$ GLSMs, Section \ref{sec4} is devoted to non-Abelian T-duality in $(2,2)$ GLSMs, Section \ref{sec5} is devoted to analyze the non-Abelian dualization of the resolved conifold GLSM, Section \ref{sec6} is devoted to $(0,2)$ GLSMs T-duality and in Section \ref{sec7} we analyze a $(0,2)$ non-Abelian dual example with $SU(2)$ symmetry. We conclude in Section \ref{sec7}, summarizing the status of GLSMs T-dualities and tracing the plan for future explorations.

\section{ T-duality in the GLSM}
\label{sec2}

In this section we describe GLSMs in 2d with $(2,2)$ SUSY and gauge group $U(1)$. We present the superfields and the Lagrangians, then we construct the Abelian T-duality procedure. Our method consists on determining the group of global symmetries of the action, and then construct a vector superfield which takes values on this group, promoting the global symmetry to be local, and adding Lagrange multipliers fields. This work has been carried out in collaboration with A.Martínez-Merino, L.Pando-Zayas and R.Santos-Silva \cite{cmps18} This is we arrive to a master Lagrangian, that upon integration of the Lagrange multiplier fields leads to the original model, and upon integration of the vector gauged fields leads to the
dual model. We present the example that reproduces MS for the $\mathbb{CP}^{N-1}$ model.

The dimensional reduction to 2d of an $\mathcal{N}=1$ 4-dimensional (4d) SUSY $U(1)$ gauge field theory gives rise to a  $\mathcal{N}=2$, $(2,2)$ theory. The $\mathcal{N}=1$ superfield language is employed to describe chiral superfields (csf)  $\Phi$ satisfiying $\bar{D}_{\dot\alpha}\Phi=0$,  antichiral superfields (acsf) $\bar{\Phi}$ satisfiying   ${D}_{\alpha}\bar\Phi=0$ and vector superfields (vsf) that fulfill the reality condition $V^\dagger=V$. The derivatives ${D}_{\alpha}$ and $\bar{D}_{\dot\alpha}$ constitute supersymmetric covariant derivatives. Twisted-chiral superfields(tcsf) satisfy the conditions  $D_+X=\bar D_-X=0$ and twisted-anti chiral superfields (tacsf) satisfy instead $\bar D_+ \bar X=D_-\bar X=0$. 

We start with the Lagrangian of a GLSM with gauge group $U(1)$ with vsf $V_0$ and $N$ csfs $\Phi_i$ with charge $Q_i$ , it can be written as \cite{witten92}
 \begin{eqnarray} 	
L_0=\int d^4\theta \left(\sum_{i=1}^N \bar\Phi_ie^{2 Q_i V_0}\Phi_i-\frac{1}{2 e^2}\bar{\Sigma}_0\Sigma_0 \right)-\frac{1}{2} \int d^2\tilde{\theta} t\Sigma_0+c.c.,\label{Lglsm}
\end{eqnarray}
where $t=r-i\theta$. The parameters of the previous Lagrangian are the $U(1)$ gauge coupling $e$, the Fayet-Iliopoulos(FI) term $r$ and the Theta angle $\theta$. These models lead to CY varieties as target space, when a superpotential depending on csf is added. In the crucial work of Witten \cite{witten92} the different phases of these models, for different values of $r$ were described.  We implement T-dualities in GLSM \cite{cmps18} by a different method than Hori and Vafa \cite{horivafa}, that for a particular Abelian case leads to the same results, i.e. MS between dual models. This
is due to the fact that the existence of a global symmetry is the origin of the duality.
Chiral superfields have the following expansion in terms of components:
\begin{align}
\Phi_i&=\phi_i(x)+\sqrt{2}\big(\theta^-\psi_i^+(x)-\theta^+\psi_i^-(x)\big)+2\theta^-\theta^+F_i(x) \label{csf}  \\
&-i\theta^-\bar{\theta}^-\partial_{03}^+\phi_i(x)
-i\theta^+\bar{\theta}^+\partial_{03}^-\phi_i(x) \nonumber \\ 
&+\frac{2i}{\sqrt{2}}\theta^+\theta^-\left(\bar{\theta}^-\partial_{03}^+\psi_i^-(x)+\bar{\theta}^+\partial_{03}^-\psi_i^+(x)\right) +\theta^+\theta^-\bar{\theta}^+\bar{\theta}^-\partial_{03}^2\phi_i(x). \nn
\end{align}
The component fields are a scalar $\phi_i$, a fermion $\psi^{\pm}_i$ and an auxiliary field $F_i$. On the other hand the vector superfield component expansion reads:

\begin{eqnarray}
V^a&&=\theta^+\bar{\theta}^+(v^a_0-v^a_3)+\theta^-\bar{\theta}^-(v^a_0+v^a_3)-\sqrt{2}\theta^+\bar{\theta}^-\sigma^a-\sqrt{2}\theta^-\bar{\theta}^+\bar{\sigma}^a+2i\bar{\theta}^+\bar{\theta}^-\notag\\
&&\times(\theta^+\lambda^{-a}-\theta^-\lambda^{+a})+2i\theta^+\theta^-(\bar{\theta}^-\bar{\lambda}^{+a}-\bar{\theta}^+\bar{\lambda}^{-a})+2\theta^-\theta^+\bar{\theta}^+\bar{\theta}^-D^a.
\end{eqnarray}
The components are the vector fields $v_{0,3}^a$, the scalar $\sigma^a$, the fermions ${\lambda^{\pm}}^a$ and the auxiliary field $D^a$. This expansion is valid for a single $U(1)$ replacing the index $a$ by $0$, as well as for multiple $U(1)$s or for the components on the adjoint representation of a non-Abelian vector superfield, where the index $a$ enumerates the generators $T_a$. The GLSM has a number of global symmetries.  Let us mention the R-symmetries. The sub-index $V$ denotes the vector R-symmetry and the sub-index $A$
denotes the axial-vector R-symmetry. The R-charges of the original fields components are given in table \ref{table1}. This is obtained by noticing that under $U(1)_V$ the transformation is: $(\theta^{\pm},\bar \theta^{\pm})\rightarrow $ $(e^{-i\alpha}\theta^{\pm},e^{i\alpha}\bar \theta^{\pm}) $ and under the $U(1)_A$ vector the transformation is: $(\theta^{\pm},\bar \theta^{\pm})\rightarrow$ $(e^{\mp i\alpha}\theta^{\pm},e^{\pm i\alpha}\bar \theta^{\pm}) $.\cite{horivafa}

\begin{table}
\small
\caption{R-charge transformations for the dual tcsf with expansion gven in (\ref{csf}).}
\begin{tabular}{|c|c|c|c|c|c|c|c|c|} 
\hline 
Field&$\phi_i$& $\psi_{i\pm}$&$F_i$&$ v^a_{0,3}$&$ \sigma^a$&$\bar\lambda^a_+$&$\lambda^a_-$&$D^a$ \\ \hline
$(q_V,q_A)$&$(q_V , 0)$&$ (q_V -1, \mp 1)$&$(q_V -1, 0)$&$(0, 0)$&$(0, 2)$&$(-1, 1)$&$(1, 1)$&$(0, 0)$\\ \hline
\end{tabular}
 \label{table1}
\end{table}

For the case of a $U(1)$ (2,2) GLSM with multiple chiral superfields $\Phi_i,\, i=1,...,N+1$  there are generically $U(1)^N$ global symmetries. If some of the charges coincide, this group of symmetries can be bigger, and in general it has the structure $U(n_1)\times U(n_2)\times ...\times U(n_k)...$, where
$n_k$ denotes a set of csfs with
equal charge. We will consider a case on which under
the $U(1)_i$ global  symmetry the csf $\Phi_i$ transform as: $U(1)_i: \Phi_i\rightarrow e^{i \hat Q_{i} \Lambda_a}\Phi_i, \, \, \, \forall_{j\neq i}\Phi_j\rightarrow \Phi_j$ and  $\Phi_{N+1}$ is uncharged under the global $U(1)$s. This is we have $N$ global $U(1)$s. Here $\Lambda_i$ is the parameter of the transformation and $\hat Q_{i}$ is the charge of the field $i$ under the transformation. The gauging of these symmetries leads to the master Lagrangian

\begin{eqnarray}
L&=&\int d^4\theta \left(\sum_{i=1}^N \bar\Phi_ie^{2 Q_i V_0+2\hat Q_i V_i}\Phi_i-\frac{1}{2 e^2}\bar{\Sigma}_0\Sigma_0+\sum_{i=1}^N(\Psi_i\Sigma_i+\bar{\Psi}_i\bar{\Sigma}_i) \right)\label{LglsmM}\\
&+&\int d^4\theta \left( \bar\Phi_{N+1}e^{2 Q_{N+1} V_0}\Phi_{N+1}\right)-\frac{1}{2} \int d^2\tilde{\theta} t\Sigma_0+c.c.,\nn
\end{eqnarray}
where Lagrange multiplier fields $\Psi_i$ have been added for every gauged  $U(1)_i$  symmetry.  The vector superfields $V_i$  constitute the gauged fields of the local $U(1)_i$ gauged symmetries. Integrating out the Lagrange multipliers one gets the pure gauge condition $\Sigma_i=0$,  including the zero vector superfield $V_i$ solution and therefore recovering the original model (\ref{Lglsm}). The original model here has target space $\mathbb{CP}^{N-1}$, for the
set of vacua with zero vev of $\phi_{N+1}$.
Therefore the duality that we are implementing should lead to the MS model, which in this case is the $A_{N-1}$ Toda variety.\cite{horivafa}

Integrating the vector superfields $V_i$ we go to the dual models. The equation of motion obtained by making the variation of the action (\ref{LglsmM}) w.r.t to $V_i$ as $\frac{\delta S}{\delta V_i}=0$  is given by
\begin{eqnarray}
\bar\Phi_i e^{2 Q_{i} V_0+2 \hat Q_i V_i}\Phi_i=\frac{\Lambda_i+\bar\Lambda_i}{2 \hat Q_i},\label{eomAbi}
\end{eqnarray}
with $\Lambda_i=\frac{1}{2}\bar D_+ D_- \Psi_i$. This integration of the $V_i$s leads to the dual Lagrangian

\begin{eqnarray}
L_{dual}&=&\int d^4 \theta\left( -\sum_{i=1}^N\frac{\Lambda_i+\bar\Lambda_i}{2 \hat Q_i}\ln \left(\frac{\Lambda_i+\bar\Lambda_i}{2 \hat Q_i}\right)+\bar\Phi_{N+1} e^{2 Q_{N+1} V_0}\Phi_{N+1}-\frac{1}{2 e^2}\bar{\Sigma}_0\Sigma_0\right) \nn
\\
&+&\frac{1}{2}\int d^2\tilde{\theta}\left(\sum_{i=1}^N \Lambda_i Q_i/\hat Q_i-t\right)\Sigma_0+\frac{1}{2}\int d^2\tilde{\bar\theta}\left(\sum_{i=1}^N \bar \Lambda_i Q_i/\hat Q_i-\bar t\right)\bar\Sigma_0.\label{LdualM}
\end{eqnarray}

Let us write now the component of the dual twisted chiral superfield. This is:


\begin{eqnarray}
\Lambda_i(\tilde X)
&=& y_i+ \sqrt{2} \theta^+ \bar\chi_{i,+} +\bar\theta^- \chi_{i,-} +2\theta^+ \bar\theta^- G_i+...\label{tcsf}
\end{eqnarray}
where we have omitted the derivatives of the component fields, represented by the three dots. The R-charges for the dual field components  are given in table \ref{table2}. The $U(1)_A$ symmetry posses an anomaly, and only a discrete parrt of it survives, which is relevant in the analysis of the dual geometry.\cite{horivafa,cs20} Let us point out that here we have obtained
the same result known for the mirror symmetric model of $\mathbb{CP}^{N-1}$,
this is, after including non-perturbative corrections what we are obtaining is the
$A_{N-1}$ Toda variety, as described
in the usual formulation of MS in GLSMs.\cite{horivafa} This can be seen because we obtain
an identical Lagrangian to the one 
in the mentioned reference. 

\begin{table}[htp]
\caption{R-charges of the dual tcsf with expansion in (\ref{tcsf}). This is obtained
by noticing the $\theta^{\pm}$ and $\bar\theta^{\pm}$ transformations under the $U(1)_V$ and $U(1)_A$.\cite{horivafa}}
\begin{center}
\begin{tabular}{|c|c|c|c|} \hline
Field&$y_i$&$ G_i$&$\chi_{i,\pm}$ \\ \hline
$(q_V,q_A)$&$(0,0)$&(0,2)&$ (-1, \mp 1)$\\ \hline
\end{tabular}
\end{center}
\label{table2}
\end{table}

To be have a more detailed description let us take as a particular example the theory with three chiral superfields $\Phi_1,\Phi_2,\Phi_3$, where 1 and 2 are equally charged, and 3 is denoted an {\it spectator field}. This denomination implies that it will not take part in the T-duality procedure.  The original GLSM has as a part of its vacua the space $\mathbb{CP}^1$. This is a nice example because it is illustrated how our procedure can be employed to obtain the mirror models. The dual Lagrangian is given by:

\begin{eqnarray}
L_{dual}&=&L_{mirror}+\int d\theta^4 \bar\Phi_{3} e^{2 Q_{0,3} V_0}\Phi_{3}.
\end{eqnarray}

This is the specialization of previous Lagrangian (\ref{LdualM}). We select the Higgs branch, determined by giving zero vevs  for the scalar components of the $U(1)$ vsf $\sigma_0$ and the spectator superfield $y_3$. The $U(1)$ field can be
integrated out giving an effective potential: 

\begin{eqnarray}
U_{eff}=2 e^2 |t - y_1 - y_2|^2. \nn  
\end{eqnarray}

In the infrared(IR) the resulting  theory is obtained by integrating the gauge field strength $\Sigma_0$ and is given by the condition $Y_1+Y_2=t$. The twisted superpotential for the non perturbative contributions in the dual theory can be written as $\widetilde W=e^{-Y_1}+e^{-t+Y_1}$.  Recall that this structure was derived by symmetry properties, and later checked to give the correct correlation functions for fermions \cite{horivafa}.

What we have obtained here is the mirror symmetric theory, originally obtained with Hori-Vafa's procedure. This is a Landau-Ginzburg model with
twisted superpotential $\widetilde W$, giving rise to the $A_1$ Toda supersymmetric field theory. Let us notice an important fact here, the full global symmetry group of the initial model is $U(1)\times U(1)$. So to obtain the mirror model, we dualize along two Abelian directions, as in Hori-Vafa procedure. But there is of course the freedom of dualizing only one $U(1)$ direction, what would have lead to a different geometry \cite{}. This is an illustration that our explorations contain MS, but they can lead to target geometries related beyond the mirror map, these explorations have been carried out by the author and collaborators \cite{cmps18}.

\section{ Non-Abelian T-duality in the GLSM}
\label{sec3}

In this section we present non-Abelian T-dualities for a $(2,2)$ SUSY $U(1)$ GLSM. The action can have  non-Abelian global symmetries; we gauge these symmetries, promoting them to be local. By integrating the new vector superfield $V$ we obtain non-Abelian T-dual models. We specify to the case of a single $SU(2)$ global symmetry. The results of this section were obtained in \cite{cmps18}.

We start presenting local non-Abelian transformations for the different superfields. The non-Abelian  global group $G$ has generators denoted by $T_a$, such that a chiral superfield $\Phi$ transforms as
\begin{eqnarray}
   \Phi'=e^{i \Lambda}\Phi,\, \, \Lambda=\Lambda^a T_a,\, \, \bar{D}_{\dot\alpha}\Lambda=0, 
\end{eqnarray}
with the transformation  $\Lambda$  satisfiying a csf condition, to preserve $\Phi$ as csf.
$\Lambda$ is in the adjoint representation of the group.  A vsf transforms under the local symmetry as $e^{V'}=e^{i\bar\Lambda} e^V e^{-i \Lambda}$, with $V=V^a T_a$.
The twisted chiral gauge field strength can be computed in terms of gauge covariant derivatives: $\Sigma=\frac{1}{2}\{\bar{\mathcal{D}}_+,\mathcal{D}_{-}\}$ and it transforms as
$\Sigma\rightarrow e^{i\Lambda}\Sigma e^{-i \Lambda}$. One obtains a master Lagrangian, by implementing the gauged symmetry and adding a Lagrange multiplier $\Psi$ in the adjoint representation of $G$, it reads in this case
\begin{eqnarray}
L_2&=&\int d^4\theta  \left( \sum_k \bar\Phi_{k,i} (e^{2 Q_k V_0+V})_{i j}\Phi_{k,j}+\tr(\Psi\Sigma)+\tr(\bar{\Psi}\bar{\Sigma})\right).\nn\\
&-&\int d^4\theta \frac{1}{2 e^2}\bar{\Sigma}_0\Sigma_0+\frac{1}{2} \left(-\int d^2\tilde{\theta} t\Sigma_0+c.c.\right).\label{eM1}
\end{eqnarray}
Integrating out $\Psi$ we obtain a pure gauge configuration, which has as a solution the original action. However there could be subtleties in this process \cite{poisson1,poisson2,poisson3}. By integrating out $V$ one goes to the dual action. The equation of motion for $V$ is given by
\begin{eqnarray} 
0&=&(D_+ \bar D_{-}\bar \Psi _a+
D_- \bar D_+ \Psi_a+\{\chi,D_+ \Psi_a\} + \{\bar\chi,\bar D_+ \bar \Psi_a\} \nonumber\\
&+&(\bar \Phi e^{2 Q V_0}e^V T_a\Phi))\Delta V_a,\label{eomV}
\end{eqnarray}
where $a$ denotes the index of the generator $T_a$,   $\chi=e^{-V}D_-e^V$ and $\Delta V= e^{-V} \delta e^V$ is a gauge invariant vector superfield variation.  The equations of motion can be simplified if one considers the following definition:
$X_a=D_+ \bar D_- \Psi_a+\{\chi,D_+ \Psi_a\} $, to give
\begin{eqnarray} 
(\bar \Phi e^{2 Q V_0}e^V T_a\Phi)=X_a+\bar X_a.\label{eomV2}
\end{eqnarray}
In general the last term in the definition is non zero, but a particular case with $\{\chi,D_+ \Psi_a\} =0$  leads to a restriction on $X_a$  giving a twisted chiral superfield (tcsf) representation. Here $\bar X_a$ constitutes an atcsf. The vsf   can be written as $V= |V| n_a  \sigma_a$, with fields $n_a$ satisfiying  the condition $\sum_a n_a^2=1$. This gives  $\tr (\Psi \Sigma)=\frac{1}{2}X_a n_a |V|$.

Let us specify to the group $G=SU(2)$, here there are three generators $a=1,2,3$, this example was developed in our first work on this subject \cite{cmps18}. Integrating out $V$, and simplifying the kinetic term of (\ref{eM1}) one gets \footnote{One has to implement a semi-chiral condition on the fields $n_a$ in order to solve the equations of motion.}:

\begin{eqnarray} 
e^{2 Q V_0} \bar \Phi_{i} e^V_{ij} \Phi_j&=& \sqrt{( X_1  + \bar X_1)^2+ (X_2+\bar X_2)^2+ (X_3 +  \bar X_3)^2},\label{eqSU2} \\
|V|&=&2 Q V_0+\ln 2 |\Phi_1|^2+\ln(\mathcal{K}(X_i,\bar X_i,n_j)).\nn
  \end{eqnarray}  
The total dual Lagrangian  can be written as
\begin{eqnarray}
L_{dual}&=&\int d^4\theta \sqrt{( X_1  + \bar X_1)^2+ (X_2+\bar X_2)^2+ (X_3 +  \bar X_3)^2}+ \label{Ldual2}\\
&+&\frac{1}{2}  \int d^4\theta  X_a n_a \ln(\mathcal{K}(X_i,\bar X_i,n_j))+c.c.\nn\\
&+&2 Q\int d\bar \theta^-d\theta^+\left(X_a n_a-\frac{t}{4}\right) \Sigma_0+c.c.\nn \\
&+& 
\frac{2 Q}{4}\int d\bar \theta^-d\theta^+\left(X_a \bar D_+n_a\right) D_-V_0+c.c.-\int d^4\theta\frac{1}{2 e^2}\bar{\Sigma}_0\Sigma_0.\nn
\end{eqnarray} 
The dual effective scalar potential is obtained by selecting the Higgs branch and integrating the $U(1)$ gauge field, it  is given by:

\begin{eqnarray}
U=2 Q^2 e^2 |x_a n_a-t/(2Q)|^2+B_a (x_a+\bar x_a),\nn
\end{eqnarray} where $x_a$ are the scalar components of the tcsf, and the $B_a$ depend on the vsf components. Fixing the gauge the new vacuum is given by the 2d space:

\begin{eqnarray}
\sum_a x_a n_a=\frac{t}{2Q}-\frac{B_1 }{2 A n_1},\,
x_2+\bar x_2=x_3+\bar x_3=0.\nn
\end{eqnarray}
The tcsf $X_{a}, a=1,2,3$ possess classical R-charges 0, as it has been summarized for the components in the corresponding table \ref{table2}, and as it can be read from the equation of motion (\ref{eqSU2}). However they present an axial anomaly\cite{horivafa,cs20}. At the quantum level there is the surviving symmetry $x_a\rightarrow x_a+\frac{2\pi i
 k_a}{2 n_a Q},\, k_a\in \mathbb{Z}$ coming from the periodicity of the parameter $t$ (the so called $\theta$ symmetry), this symmetry can be checked already in the dual Lagrangian (\ref{Ldual2}). This symmetry is crucial at the time of analyzing the dual geometry. The vacua gives as a dual target space the 2d torus $T^2$.

For an Abelian direction inside $SU(2)$ with $n_a=const$  the non-perturbative corrections on the dual theory are identified as $\tilde W=e^{-X_a n_a}$, these play the same role that the instanton vortex solutions on the dual model \cite{horivafa}. The original and the dual models posses the same effective superpotential for the $U(1)$ gauge field; this constitutes an important match of both theories.  For a pure non-Abelian direction we still have to obtain what are the non-perturbative corrections to the superpotential.

\section{ Duality in a non compact CY}
\label{sec4}

Let us now consider a $(2,2)$ SUSY GLSM with $G=SU(2)\times SU(2)$ global symmetry. We will review the dualization procedure for this model, as well we will discuss the geometry.  First we dualize an Abelian direction inside $G$, then we will consider a non-Abelian dualization. The results presented here are based on the work in collaboration with Y.Jimenez-Santana and R-Santos-Silva\cite{cjs22}. We add new content on the discussion of the periods of the dual geometry.

The model consists of two pairs of csf with equal charge under $U(1)$ gauge symmetry. Therefore one has $SU(2)\times SU(2)$ global symmetry. Gauging the global symmetry gives rise to the master Lagrangian:
\begin{eqnarray}
L_{model}&=&\int d^4\theta  \left(  \bar\Phi_i (e^{2 Q V_0+V_1})_{i j}\Phi_{j}+\bar\Phi_l (e^{2 Q V_0+V_2})_{lm}\Phi_{m} \right)\label{LMSU2} \\
&+&\int d^4\theta \left( \tr(\Psi_1\Sigma_1)+\tr(\Psi_2\Sigma_2)+c.c \right)\nonumber\\
&-&\int d^4\theta  \frac{1}{2 e^2}\bar{\Sigma}_0\Sigma_0 -\frac{1}{2} \left(\int d^2\tilde{\theta} t\Sigma_0+c.c.\right), \nn \label{nablsu2}
\end{eqnarray}
with $i,j=1,2$ and $l,m=1,2$, indices for $SU(2)$ doublets for chiral and anti-chiral superfields. Integrating out the Lagrange multipliers $\Psi_1,\Psi_2$ the original action is obtained. While by integrating the vector superfields $V_1$ and $V_2$ one obtains the dual model.

The original action consists of  two csf $\Phi_1,\Phi_2$ with charge 1 and  two csf $\Phi_3,\Phi_4$ with charge -1 under the $U(1)$. The locus of the zero scalar potential reads: \begin{eqnarray}
|\phi_1|^2+|\phi_2|^2-|\phi_3|^2-|\phi_4|^2=t.
\end{eqnarray} The $\phi_i$ represent the scalar components of the $\Phi_i$ csf. By modding out the $U(1)$ symmetry the resulting geometry is the one of the resolved conifold.  Integrating in (\ref{LMSU2}) the gauged vector fields $V_1$ and $V_2$ the dual model gives two copies of the same Lagrangian \begin{eqnarray}
L_{dual,0}&=&\int d^4\theta \left(\sqrt{\sum_a( X_a  + \bar X_a)^2}+ \sqrt{\sum_a( Y_a + \bar Y_a)^2}\right)\label{exT}\\
&+& \int d^4\theta  (X_a \hat n_a \ln(\mathcal{K}(X_i,\bar X_i,n_j))+X_a \hat m_a \ln(\mathcal{K}(Y_i,\bar Y_i,n_j)))+c.c.\nn\\
&+&\frac{1}{2}\int d\bar \theta^-d\theta^+\left(2 Q X_a n_a-2 Q Y_am_a-t\right) \Sigma_0+c.c.,\nn\\
&+&\frac{1}{2}\int d\bar \theta^-d\theta^+\left(Q X_a \bar D_+n_a-Q Y_a\bar D_+m_a\right) D_-V_0+c.c.\nn \\
&-& \int d^4\theta\frac{1}{2 e^2}\bar{\Sigma}_0\Sigma_0 -\frac{1}{2} \left(\int d^2\tilde{\theta} t\Sigma_0+c.c.\right).\nn
\end{eqnarray}
Here as in Mirror symmetry: The csf $\Phi_i$ are exchanged by  twisted csf $X_a$ and $Y_a$. The charges can be selected to be $Q=1$.  The Langrangian has the quantum symmetry 

                              \begin{eqnarray}
 Im(X_{a}) \rightarrow  Im(X_{a}) + \frac{\pi i k^{Ia}}{Q}, \, k^{Ia}\in \mathbb{Z},
     \nonumber\\
      Im(Y_{a}) \rightarrow  Im(Y_{a}) + \frac{\pi i k^{IIa}}{Q}, \, k^{IIa}\in \mathbb{Z},
     \nonumber
         \end{eqnarray}coming  from the $t$ periodicity, that can be observed from the dual action (\ref{exT}). It gives rise to the scalar potential \begin{eqnarray}
U=8 e^2 (t + 
   Q (-n_{1} X_{1} - n_{2}X_{2} - n_{3} X_{3} + m_{1} Y_2 + m_{2} Y_2 +    m_3 Y_3))\times  \nn \\
(\bar{t} + 
   Q (-n_{1} \bar X_{1} - n_{2}\bar X_{2} - n_{3} \bar X_{3} + m_{1} \bar Y_2 + m_{2} \bar Y_2 +    m_3 \bar Y_3)). \nn
      \end{eqnarray}
One has the gauge freedom to fix the gauge to be
            \begin{eqnarray}
X_{2}+\bar X_{2} =X_{3} +\bar X_{3} =Y_2 +\bar Y_2 = Y_3+\bar Y_3=0. \nn
      \end{eqnarray}
   The SUSY vacuum  is given by the  set of six real equations:
      \begin{eqnarray}
    && Q (-n_{1} Re(X_{1}) - n_{2}Re(X_{2}) )=-t_1. \\
     &&Q (-n_{1} Im(X_{1}) - n_{2}Im(X_{2}) - n_{3} Im(X_{3}) )
\\&+& Q( m_{1}Im( Y_2) + m_{2} Im(Y_2) +    m_3 Im(Y_3))=-t_2.
     \nonumber
         \end{eqnarray} 
At a perturbative level the dual model geometry is given by: $T^5\times \mathbb{R}$.  But non perturbative effects as instanton corrections have to be considered in the dual model, for the case of $n_a=const$, because in the original model has vortex solutions \cite{horivafa}.

One important check is to compute the effective twisted superpotential for the $U(1)$ gauge field in both models. In the original model one obtains

\begin{eqnarray}
 W_{eff}(\Sigma_0)=\sum_i -\Sigma_0 Q_i\left(\ln\left(\frac{Q_i\Sigma_0}{\mu}\right)-1\right)-t \Sigma_0.
\end{eqnarray}
This serves to propose an Ansatz for the non perturbative effects, corresponding to the original model instanton corrections: $\widetilde W_{np}=(e^{-X_a n_a}+e^{-Y_am_a})$; those are included in the dual Lagrangian as:
 \begin{eqnarray}
L_{dual}&=&L_{dual,0}+ 2\mu \int d^2\tilde\theta(e^{-X_a n_a}+e^{-Y_am_a})+c.c. \nn
  \end{eqnarray}
This proposal is valid for an Abelian direction inside the non-Abelian duality. This twisted superpotential in the dual theory gives the correct effective superpotential for the $U(1)$ vector field,  and the dual geometry is still: $T^5\times \mathbb{R}$.  For the case of non-Abelian duality we still have to investigate the appropriate non-perturbative corrections.

Moreover a trivial deformation on the infrared can be added to give another model, by following the approach presented in \cite{HoriTrieste}. The motivation is to obtain a geometry that has the same number of dimensions, in this case 3 complex dimensions. Adding an irrelevant quadaratic change on the twisted superpotential $\Delta \widetilde W=u^2$, with a new field $u$ we obtain the locus:
\begin{eqnarray} 
\widetilde W=x_1 x_2 x_3 \left(1 + e^{-t/2}\right)+ u^2=0. \label{dW}
\end{eqnarray}
The change of variables $x_a=e^{-X_a n_a}$ (without summation) is in place. This change of variables maps the periodic
variables  $X_a$ to the space  $(\mathbb{C}^\times)^3$.  This zero locus constitutes a threefold geometry. This is a cubic polynomial in $(\mathbb{C}^\times)^3 \times \mathbb{C}$ with the coordinates $(x_1,x_2,x_3,u)$. Notice that for $e^{-t/2}=-1$ we have a singularity, which is satisfied by $t/2=(2k+1)\pi i, k\in \mathbb{Z}$. The set of critical points is the whole space $(\mathbb{C}^{\times})^3$ for the coordinates  $(x_1,x_2,x_3)$. This can also be seen from (\ref{dW}), when the first term is zero the solution  is $u=0$. For this case $r=0$ the original theory is the singular conifold. Thus both models show a singularity for the same value of the parameters.

The match between the original and dual theory has as a first observable the matching of this $U(1)$ field effective potential, additionally one can read
that in the parameter $t$ also posses a singularity at the conifold point in the original model. Is as well possible to compute the periods in this dual geometry employing the prescription of Hori and Vafa.
The period integral reads:
\begin{eqnarray}
 \Pi&=&\int d X_1 dX_2 dX_3 d Y_1 dY_2 d Y_3 \delta(2 X_a n_a-2 Y_a m_a-t)exp\left(-W \right)
 \end{eqnarray}
We eliminate the variable $Y_3$ by integrating:

\begin{eqnarray}
 \Pi&=&-\frac{1}{2m_3}\int d X_1 dX_2 dX_3 d Y_1 dY_2 exp\left(-W \right)\\
 W&=&e^{-X_a n_a}+e^{-X_a n_a+t/2}. \nonumber
 \end{eqnarray}
The periods read:
 \begin{eqnarray}
 \Pi&=&-\frac{1}{n_1 n_2 n_3}\left(\int d Y_1 dY_2\right)\int \frac{d x_1 dx_2 dx_3}{x_1 x_2 x_3}  exp\left(-(1+e^{t/2}){x_1} {x_2}{x_3} \right).
 \end{eqnarray}
We can take the cycle $S_1\times S_1\times S_1$, to obtain a period  $\Pi_a=const$.

If we make the derivative of the period integral with respect to $t$, one gets:
\begin{eqnarray}
 \frac{\partial \Pi}{\partial t}&=&\frac{e^{t/2}}{n_1 n_2 n_3}\left(\int d Y_1 dY_2\right)\int d x_1 dx_2 dx_3 exp\left(-(1+e^{t/2}){x_1} {x_2}{x_3} \right)\\
 &=&\frac{e^{t/2}}{n_1 n_2 n_3 (1+e^{t/2})}\left(\int d Y_1 dY_2\right)\int \frac{dx_2 dx_3 }{x_2 x_3} \times \\
 &\times&\int d ((1+e^{t/2})x_1 x_2 x_3 )exp\left(-(1+e^{t/2}){x_1} {x_2}{x_3} \right). \nonumber
 \end{eqnarray}
If we can now have $x_2$ and $x_3$ in $S^1\times S^1$, and $v=(1+e^{t/2})x_1 x_2 x_3$ from $0$ to $\infty$ then we have: \begin{eqnarray}
 \frac{\partial \Pi_b}{\partial t}&\sim& \frac{e^{t/2}}{(1+e^{t/2})},\\
\Pi_b &\sim& \log(1+e^{t/2})= \log(1+\frac{1}{\sqrt{z}})\\
&=& -Li_1(\frac{-1}{\sqrt{z}}),\nonumber
 \end{eqnarray} with $z=e^{-t}$. Thus we encounter two period solutions for the dual geometry, the constant and $-Li_1(\frac{-1}{\sqrt{z}})$. The result is the same for the deformation by an irrelevant term in the IR as in equation (\ref{dW}). These periods can be compared with those ones in a more recent version of\cite{cjs22}.

\section{ T-Dualities in 2d (0,2) GLSMs }
 \label{sec5}

 In this section we will review more recent work, carried out in collaboration with R.Díaz-Correa and H.García-Compeán\cite{cdg23}. This is the formulation of the current T-duality procedure in GLSMs with $(0,2)$ SUSY. We consider here $(0,2)$ models which come from a $(2,2)$ reduction or pure $(2,0)$ models. 
 
 Let us analyze the first case. We start with a Lagrangian with  $n$ chiral superfields $\Phi_i$, $\tilde{n}$ Fermi superfields $\Gamma_i$ and $m$ $U(1)_a$, $a=1...m$, gauge symmetries with vector superfield components $V_a,\Psi_a$ and field strength $\Upsilon_a$:
\begin{eqnarray}
\label{genCase}
L&=&\I\bigg\{\sum_{a=1}^m \frac{1}{8e_a^2}\bar\Upsilon_a\Upsilon_a-\sum_{i=1}^n\frac i2\overline\Phi_i e^{2 \sum_{a=1}^m Q_i^a\Psi_a}\bigg(\partial_{--}+i\sum_{a=1}^m Q_i^aV_a\bigg)\Phi_i \nonumber \\
&+&\I\bigg\{\sum_{i=1}^n\frac i2\overline\Phi_i(\overleftarrow{\partial}_{--}-i \sum_{a=1}^m Q_i^aV_a)e^{2\sum_{a=1}^m Q_i^a\Psi_a}\Phi_i\bigg\}  \nonumber\\
&-& \I\bigg\{\ \sum_{i=1}^{\tilde{n}} \frac12e^{2 \sum_a \tilde{Q}_i^a\Psi_a}\overline\Gamma_i\Gamma_i\bigg\} \nonumber  \\
&+&\sum_a \frac{t_a}{4}\int\text d\theta^+\Upsilon_a|_{\bt=0}.  \nonumber
\end{eqnarray}

The fields $\Phi_i$ and $\Gamma_i$ have charges $Q_i^a$ and $\tilde Q_i^a$ under the $U(1)_a$ gauge symmetry group. Let us further assume that there are global symmetries; those are generically $n-m$ $U(1)$s for the chiral superfield sector, and $\tilde n-m$ $U(1)$s for the Fermi superfield sector. Let us gauge the global symmetries of the chiral sector, obtaining an associated field strength $\Upsilon_b$ and adding Lagrange multipliers $\Lambda_b$  for each one, this
leads to the master Lagrangian:

\begin{eqnarray}
L&=&\I\bigg\{\sum_{a=1}^m \frac{1}{8e_a^2}\bar\Upsilon_a\Upsilon_a \nn \\
&-&\I\sum_{i=1}^k\frac i2\overline\Phi_i e^{2 \sum_{a=1}^{m} Q_i^a\Psi_a+2 \sum_{b=1}^{\tilde{k}} Q_{1i}^b\Psi_{1b}}\bigg(\partial_{--}+i\sum_{a=1}^m Q_i^aV_a+i\sum_{b=1}^k Q_{1i}^bV_{1b}\bigg)\Phi_i  \nonumber \\
&+&\I\bigg\{\sum_{i=1}^k\frac i2\overline\Phi_i \bigg(\overleftarrow{\partial}_{--}-i \sum_{a=1}^m Q_i^aV_a-i \sum_{a=1}^m Q_{1i}^bV_{1b}\bigg)e^{2\sum_a {Q_i^a\Psi_a+Q_{1i}^b\Psi_{1b}}}\Phi_i\bigg\}  \nn \\
&-& \I\bigg\{\ \sum_{i=1}^{\tilde{n}} \frac12e^{2 \sum_a \tilde{Q}_i^a\Psi_a+2 \sum_c \tilde{Q}_{1i}^c\Psi_{1c}}\overline\Gamma_i\Gamma_i\bigg\} \nonumber  \\
&+&\sum_a \frac{t_a}{4}\int\text d\theta^+\Upsilon_a|_{\bt=0}+\sum_{b=1}^{k}\I \Lambda_b \Upsilon_b+\sum_{c=k+1}^{k+\tilde{n}}\I \Lambda_c \Upsilon_c+c.c.\nonumber\\
&-& \frac i2 \I \sum_{i=k+1}^n \bigg\{\overline\Phi_i e^{2 \sum_a Q_i^a\Psi_a}(\partial_{--}+i\sum_a Q_i^aV_a)\Phi_i  \nonumber\\
&-&\overline\Phi_i \bigg(\overleftarrow{\partial}_{--}-i \sum_a Q_i^aV_a\bigg)e^{2\sum_a Q_i^a\Psi_a}\Phi_i\bigg\}.   \nonumber
\end{eqnarray}

Integrating the Lagrange multipliers, one goes to pure gauge configurations, that allow to recover the original model. Performing the variation of the Lagrangian with respect to the vector superfield component $\Psi_{1b}$ we obtain:

\begin{eqnarray}
V_{1b}&=&A^{-1}_{bd}(-\frac{i}{2}\partial_-Y_-^d-R^d), \\
A_{bd}&=&\sum_{i=1}^k |\phi_i|^2Q_{1i}^dQ_{1i}^b, \nn \\
R^d&=&\sum_{i=1}^k \left(-\frac{i}{2}\bar \Phi_i \delta_- \Phi_i Q_{1i}^d+|\Phi_i|^2\sum_{a=1}^
m Q_i^a V_a Q_{1i}^d\right), \nn
\end{eqnarray}
with the dual field $Y_\pm^c=i\bar D_+\Lambda^c\pm iD_+\bar\Lambda^c$; and with the definition $\delta_-=\partial_--\overleftarrow\partial_-$. Notice as well the relevant definitions $R^d$ and $A_{bd}$, which are paremeters depending on the original chiral superfields, and have to be fixed by the gauge freedom. The vector superfield components $V_{1b}$ are determined in function of those parameters. Performing the variation of the Lagrangian with respect to $V_{1b}$  one has:
\begin{eqnarray}
\psi_{1b}&=&A^{-1}_{bd}(-\frac{i}{2}\partial_-Y_+^d-S^d), \\
A_{bd}&=&\sum_{i=1}^k |\phi_i|^2Q_{1i}^dQ_{1i}^b, \nn \\
S^d&=&\sum_{i=1}^k |\Phi_i|^2 Q_{1i}^d+2 \sum_{a=1}^m |\Phi_i|^2 Q_{1i}^d Q_{1i}^a\psi_a. \nn
\end{eqnarray}
Notice that now the vector superfield components $\psi_{1b}$ are given in terms of the parameters $A_{bd}$ and $S^d$, which also need to be fixed by the gauge freedom.
For the component $\psi_{1d}$ we have:
\begin{eqnarray}
 Q_{1 j}^d\bar\Gamma_j \Gamma_j=-i \partial_- Y_-^d \rightarrow \bar\Gamma_j \Gamma_j=-Q^{-1}_{1jd}\partial_-Y_-^d. \nn
\end{eqnarray}
The dual models can be obtained by solving these equations of motion
for the vector superfield.
\subsection*{Building block example}

In the following we start from a simple $(0,2)$ model coming from a $(2,2)$ reduction, with a  single chiral superfield, and a single Fermi superfield of equal charges.  This duality is the building block of the mirror map described in \cite{abs03}. The original model gives rise to a point with $|\phi|^2=r$ and $E_0=J_0=0$. One can gauge the global  $U(1)$  symmetry to  obtain the Lagrangian:

\begin{eqnarray}
L&=&\I \bigg\{-\frac i2\overline\Phi e^{2(\Psi_0+\Psi_1)}\big[\partial_{--}+i(V_0+V_1)\big]\Phi  \\
&+&\frac i2\overline\Phi\big[\overleftarrow{\partial}_{--}-i(V_0+V_1)\big]e^{2(\Psi_0+\Psi_1)}\Phi +\I\frac{1}{8e^2}\bar\Upsilon_0\Upsilon_0 \nn \\
&-&\frac12e^{2(\Psi_0+\Psi_1)}\overline\Gamma\Gamma+\Lambda\Upsilon_1+\overline\Upsilon_1\overline\Lambda+\overline\chi\widetilde E_1+\overline{\widetilde{E}}_1\chi\bigg\}\;+
\frac t4\int\text d\theta^+\Upsilon_0|_{\bt=0}+ {\rm h.c.}+... \nn
\end{eqnarray}
Integrating the Lagrange multipliers $\Lambda$ and $\chi$ one obtains the original model.  Integrating the gauge fields $V_1$ and $\Psi_1$, the dual Lagrangian:
\begin{eqnarray}
L_{dual}&=&\I\bigg\{-\frac i2\frac{Y_-\partial_{--}Y_+}{Y_+}+\frac{|\Phi|^2\F\overline\F}{Y_+}-(\Lambda\Upsilon_0+\overline\Upsilon_0+\chi E_0+\overline E_0\overline\chi)\bigg\} \nn \\
&+&\frac t4\int\text d\theta^+\Upsilon_0|_{\bt=0}+ {\rm h.c.}+\I\frac{1}{8e^2}\bar\Upsilon_0\Upsilon_0. \nn
\end{eqnarray}
This is the building block of the description of mirror symmetry for $(0,2)$ models arising from a supersymmetric reduction as discussed by Adams, Basu and Sethi.\cite{abs03}

\section{Non-Abelian T-duality of 2d  $(0,2)$ GLSM}

\label{sec6}

In this section we construct non-Abelian dual 2d $(0,2)$ models.\cite{cdg23} This is part of a recent collaboration work \cite{cdg23}. We obtain the dual Lagrangian, and specify for the global symmetry group $SU(2)$, finding the dual model.

Let us start by a master Lagrangian, which has the gauged symmetry $G=SU(n_1)\times...\times...SU(n_s)$
and Lagrange multipliers $\Lambda_I$:

\begin{eqnarray}
L&=&\I\bigg\{\sum_{a=1}^m \frac{1}{8e_a^2}\bar\Upsilon_a\Upsilon_a\\
&-&\I\sum_{I=1}^s\frac i2\overline\Phi_I e^{2 \sum_{a=1}^{m} Q_I^a\Psi_a+2 \Psi_{1I}}(\partial_{--}+i\sum_{a=1}^m Q_I^aV_a+i V_{1I})\Phi_I  \nonumber \\
&+&\I\bigg\{\sum_{I=1}^s\frac i2\overline\Phi_I(\overleftarrow{\partial}_{--}-i \sum_a Q_I^aV_a-i V_{1I})e^{2\sum_a {Q_I^a\Psi_a+2 \Psi_{1I}}}\Phi_I\bigg\}  \nn \\
&-& \I\bigg\{\ \sum_{I=1}^{s} \frac12 (\overline \Gamma_I+\overline \Gamma_{1I}) e^{2 \sum_a Q_I^a\Psi_a+2 \Psi_{1I}}(\Gamma_I+\Gamma_{1I})\bigg\} \nonumber  \\
&+&\sum_a \frac{t_a}{4}\int\text d\theta^+\Upsilon_a|_{\bt=0}+\sum_{I=1}^{s}\I  \Tr \{\Lambda_I \Upsilon_I\}+c.c. \nn \\
&+&\sum_{I=1}^{s}\I   \bar{\chi}_I  \BLE_I+c.c.\nonumber
\end{eqnarray}
This master Lagrangian has been obtained by gauging global symmetries in a $(0,2)$ $U(1)^n$ GLSM with chiral and Fermi superfields. The $\Phi_I=(\Phi_{I1},...\Phi_{In_I})$  are vectors of chiral superfields, with $I=1...s$. $n_I$ denotes
the number of chiral superfields with equal charges under $U(1)^n$.
$V_{1I}=V_{Ib}T_b$, $\Psi_{1I}=\Psi_{Ib}T_b$ are vector superfields for each gauged group $SU(n_I)$. The gauging of global non-Abelian symmetries has been performed on a model coming from a reduction of a $(2,2)$ SUSY; where the Fermi fields coincide in number with the chiral fields and have coincident charges. In principle the group of global symmetries of the original model is bigger,
with the structure:  $\tilde G=U(n_1)\times...\times...U(n_s)$, but we select only the special sector of the unitary groups. These wider duality group can
be explored elsewhere.\footnote{We thank David Tong for pointing out this important difference for the $(2,2)$ SUSY case.} The original model is obtained by obtaining the equations of motion for the Lagrange multipliers $\Lambda_I$ and $\bar\chi_I$.

On the other hand, if we want to obtain the dual model, we have to integrate the vector superfield components. The dual fields are: $\F_I=e^{\Psi_I}\D_+\chi_I$ and the following definitions are given: $a^{ab}_I:=\BLPhi_I^\dagger\{\T^a,\T^b\}\BLPhi_I$, $e_I:=1_I+2\sum_{\alpha=1}^mQ_I^\alpha\Psi_\alpha$, $Z_I^a:=\BLPhi_I^\dagger\T^a\BLPhi_I$. 
The variations with respect to gauge fields $V_{1I}^c$, $\Psi_{1I}^c$ and $\Gamma_{1I}^c$ give:
\begin{eqnarray}
\delta_{V_{1I}^c}S&=&0:\qquad\qquad\Psi_{1I}^aa^{ca}=-Y_{+Ia}\Tr(\T^a\T^c)-e_IZ_I^c:=K_I^c \nn\\
\delta_{\Psi_{1I}^c}S&=&0:\quad V_{1I}^ba^{bc}_I+2Q_I^\beta V_\beta Z_I^c-i\Phi_I^\dagger\T^c\delta_-\Phi_I+i\partial_-Y_{-aI}\Tr(\T^a\T^c)\nn \\
&-&(\Gamma^\dagger_I+\Gamma_I^{a\dagger}\T^a)\T^b(\Gamma_I+\Gamma_I^c\T^c)=0 \nn \\
\delta_{\Gamma_{1I}^c}S&=&0:\qquad\qquad-\frac12(\Gamma_I^\dagger+\Gamma_{1I}^{a\dagger}\T^a)(e_I+2\Psi_{1I}^b\T^b)\T^c-\frac{\sqrt2}{2}\F_I^{a\dagger}\T^a\T^c=0. \nn
\end{eqnarray}

Thus for the group $\mathfrak{g}=\mathfrak{g}_1\times\mathfrak{g}_2\times\cdots\times\mathfrak{g}_s$ the dual Lagrangian becomes:

\begin{eqnarray}
L_{dual}&=&\sum_{I=1}^s\I\bigg\{-\frac i2e_I\BLPhi_I^\dagger\delta_-\BLPhi_I+\BLPhi^\dagger_I\BLPhi_Ie_IQ_I^\beta V_\beta+\BLF_I^\dagger X_I^{-1}\BLF_I \nn \\
&+&\frac{\sqrt2}{2}(\BLF_I^\dagger\BLGamma_I+\BLGamma_I^\dagger\BLF_I)\\
&\times&[-i\BLPhi_I^\dagger\delta_-\T^a\BLPhi_I+2Q_I^\beta V_\beta Z_I^a][-Y_{+Ib}\Tr(\T^b\T^c)-e_IZ_I^c]b_{ac}\bigg\}, \nn
\end{eqnarray}
where we have the definitions $X_I:=e_I+2\T^aK_I^a=e_I-2\T^ae_IZ_I^cb^{ca}-2\T^aY_{+b}\Tr(\T^b\T^c)b^{ca}$
and $b^{ac}={(a^{ac})}^{-1}$.
One has to perform gauge fixing to remove the original chiral fields $\BLPhi$. Next we will discuss a simple example, explored in the work summarized in this section \cite{cdg23}.

\subsection*{ Example of an $SU(2)$ dual}

Now let us look at a dualization example. Consider a $(0,2)$ U(1) GLSM obtained from reduction of a $(2,2)$ U(1) GLSM
with two chiral fields and two Fermi fields, such that $E=iQ\sqrt2\Sigma'\Phi'$
and $\Sigma'=\Sigma|_{\theta^-=\overline\theta^-=0}$, and $\Phi'=\Phi|_{\theta^-=\overline\theta^-=0}$ \cite{abs03}. Gauging the global  $SU(2)$  symmetry one can obtain the dual Lagrangian:

\begin{eqnarray}
L_{dual}&=&\sum_{I=1}^s\I\bigg\{Q_\beta V^\beta\Big[e_I-e_I\frac{Z^aZ_a}{Z_0}-\frac{Y_+^aZ^a}{Z_0}\Big]\label{Lsu2} \\&+&\BLF^\dagger \Big(e_I\mathbf{I_d}-\frac{\T^a}{Z_0}(e_IZ^a_I+2Y_+^a)\Big)^{-1}\BLF\nn\\
&+&\frac{\sqrt2}{2}(\BLF_I^\dagger\BLGamma_I+\BLGamma_I^\dagger\BLF_I)\bigg\}+ \frac{t}{4}\int\text d\theta^+\Upsilon|_{\bt=0}, \nn
\end{eqnarray}
we have the definitions $e_I:=1_I+2\sum_{\alpha=1}^mQ_I^\alpha\Psi_\alpha$, $X_I:=e_I\mathbf{I_d}-\T^a\frac{e_IZ^a_I+2Y_+^a}{|\Phi_I|^2}$, $Z_I^a:=\BLPhi_I^\dagger\T^a\BLPhi_I$ and $\BLF_I$. The original model gives rise to the $\mathbb{CP}^1$ target space with:
$|\phi_1|^2+|\phi_2|^2=r$ and $E_0=J_0=0$. The dual model in (\ref{Lsu2}) has target given by the surface:
\begin{equation}
\frac e2(\Im(t)-y_+^aZ_a)^2+(\bar E_1\quad\bar E_2)A\begin{pmatrix}E_1\\E_2\end{pmatrix}=0,\label{su2Dual}
\end{equation}
with  the matrix $A=\frac{-2y_+^by_+^b}{1-u_cu^c}\begin{pmatrix}u_3-1&u_{12}\\\bar u_{12}&-1-u_3\end{pmatrix}$. We have 
$u^a=2\frac{y_+^a}{Z_0}+\frac{Z_a}{Z_0}$, with $Z_0=\overline\Phi_1\Phi_1+\overline\Phi_2\Phi_2$. We take a gauge with the $Z_a=const$ in particular $Z_0=1$.

An analysis of the geometry leads to a disk for the case of  a positive definite scalar potential\cite{cdg23}.   As we have three real variables $y^1_+,y^2_+,y^3_+$, then the vacua consist of a two-dimensional surface. For a semi-definite positive potential and the surface $y_+^a Z_a=\Im(t)$  we have a plane inside the sphere, this is a disk $\bf D$. The parameter $r=\Im(t)$ determines the size of the disk and its position. Also if $r\notin[-1,0]$ this disk is empty. For the case outside the sphere one has a surface given by the locus (\ref{su2Dual}).

Non perturbative corrections to the superpotential, which match the original model instanton contributions, can be argued to have the form: $W= \beta  e^{\alpha^b Y_b}$. These can be derived from the reduction of $(2,2)$ SUSY. One can also obatin them in the pure $(0,2)$ model by following the argument of Hori and Vafa \cite{horivafa} to determine the shape of the Ansatz of non-perturbative twisted-superpotential contributions. This is an argument based on: holomorphicity in $t$, periodicity in $\theta$, R-symmetry and the asymptotic behaviour. One important point is that the superpotential has axial R-symmetry with charge 2.
When these non-perturbative corrections are included, the only change of the model of this subsection is a shift of the holomorphic function ${E}$. Therefore we also have the same dual geometry as in the previous case. Obtaining that for the $\mathbb{CP}^1$ 2d $(0,2)$ GLSM the dual model target space is a disk $\bf D$\cite{cdg23} .

\section{Discussion and conclusions} 
\label{sec7}

In this work we review a method to construct T-duality  in supersymmetric 2d GLSMs, for the Abelian and the non-Abelian symmetries \cite{cmps18,cs20,cjs22,cdg23}. First we have studied the $(2,2)$ SUSY case \cite{cmps18} and more recently
the $(0,2)$ SUSY case \cite{cdg23}.  For $(2,2)$ SUSY, the Abelian T-duality for models with CICY target spaces gives rise to the  mirror symmetric models studied in the seminal work of Hori and Vafa \cite{horivafa}.  As well for  $(0,2)$ SUSY the Abelian T-duality gives rise to the mirror models obtained as reduction by Adams, Basu and Sethi \cite{abs03} . Therefore our procedure is able to recast mirror symmetry for the case of Abelian T-duality, for an specific choice of charges. However our procedure also casts a wider group of Abelian T-dualities. 

Furthermore, our techniques serve to investigate correspondences beyond MS. We have developed the construction of non-Abelian T-dualities in GLSMs, obtaining dual geometries, and so far we have analyzed various examples.   The first example of T-duality computed  for the non-Abelian case was the dual of the $(2,2)$ GLSM with $SU(2)$ global symmetry, which original model leads to the $\mathbb{CP}^1$ model. Here we observe, that
the duality for an Abelian subgroup inside $SU(2)$ gives the dual geometry of $T^2$, with and without non-perturbative corrections. Furthermore a truly non-Abelian simplified version of the duality, without non-perturbative corrections leads also for the dual geometry $T^2$. This is the first example where we observe this fact.

We have reviewed as well the implementation of non-Abelian T-duality for $(2,2)$ 2d U(1) GLSMs  with with $\rm SU(2)\times SU(2)$ global symmetry, this is the first example of a non-compact CY manifold \cite{cjs22}. In a simplified version for the duality, for an Abelian direction inside $\rm SU(2)\times SU(2)$,  without non perturbative terms one obtains as the dual geometry: $T^5\times \mathbb{R}$. Incorporating non perturbative corrections  the same geometry is preserved, as in the previous case of a single $SU(2)$. When we consider a truly non-Abelian duality, it also leads to the same geometry.  To match the dimensionality, of the dual model, we  have added a deformation to the superpotential irrelevant in the IR \cite{HoriTrieste}, to obtain a three-fold which corresponds to the dual geometry of the resolved conifold. The new content here, is that we compute the periods on the  dual model \cite{horivafa}, obtaining two independent solutions. This is done for the dual model without deformation and it coincides with the period calculation for the dual model with the deformation, which can be found in a recent version of the work \cite{cjs22}. Let us emphasize that the correspondence, for the case of a direction inside the non-Abelian group,  is supported by the calculation of the effective potential for the $U(1)$ gauge field, which matches in both models, once that non-perturbative contributions are taken into account. Additionally the $t$ parameter possess
a singularity in the original and in the dual description \cite{cjs22}.

We review the extension of the method to describe T-duality, Abelian and non-Abelian, in supersymmetric $(2,0)$ 2D GLSMs for models coming from a reduction of the $(2,2)$ case and for the general case, developed in \cite{cdg23}. As an example we analyze the case of $SU(2)$ global symmetry, where the original geometry is $\mathbb{CP}^1$, obtaining as a dual geometry a two dimensional surface, which for the positive definite potential leads to a disk. The main advance of this investigations with respect to the $(2,2)$ case is that the we don't require to specify an Abelian direction inside the duality group. Here it is possible to solve the equations of motion fully. An important observation is
that by considering non-perturbative corrections in the dual models, the geometry of the duals is preserved. This coincides with what we have observed for the $(2,2)$ dualities.

The main goal of the work is to explore the whole group of T-dualities in GLSMs, and to identify to which geometric correspondences these may lead. These could be connected to Mirror Symmetry, or extensions of it. For this purpose there are aspects to be explored, which we summarize next:
\begin{itemize}
    \item 
For $(2,2)$ SUSY models we need to solve the total non-Abelian T-duality. As mentioned, in contrast, for  $(0,2)$ SUSY models we were able to solve the equations of motion generically. One direction planned, is to consider the equations of motion leading to the dual $(2,2)$ Non-Abelian model for any gauged group, working in the components language, instead of the superfield language described here. 

\item In the $(2,2)$ case the non-perturbative effects  in the full non-Abelian dual models have to be studied.
Currently, these corrections are established for an Abelian direction inside the non-Abelian group.

\item As well for $(2,2)$ SUSY it would be relevant  to compute the partition function on $S^2$ of the non-Abelian dual models with localization techniques, to compare them with the partition function of the original GLSMs \cite{localization1,localization2,localization3,localization4} putting the duality in solid grounds \cite{cmps24}.

\item The geometrical interpretation for dual target spaces of the non-Abelian T-duality is a fundamental question in our work. In particular we plan to study the connection between the non-Abelian T-duality in the GLSMs to MS in examples of CY varieties, possibly determinantal varieties \cite{determinantal,gusharpe5}.  There are already descriptions of MS for these models, with which it would be relevant to compare \cite{gusharpe1,gusharpe2,gusharpe3,gusharpe4}.

\item We also plan to explore target compact manifolds considering a superpotential in the original theory for both SUSY cases. 
\item We would like as well to extend the duality to other $(2,2)$ SUSY representations as semi-chiral superfields, which give rise to geometries with torsion. 

\item We also plan to develop the non-Abelian T-dualities described here in the context of Non-Abelian GLSMs explored by Hori and Tong \cite{horitong}, where there could be interesting dynamics arising. 

\end{itemize}

To summarize, the main goal of the explorations presented here, is to obtain new geometric and physical correspondences for string theory on CY manifolds.

\section*{Acknowledgments}

 I  would like to specially thank: Jorge Gabriel León Bonilla, Rodrigo Díaz Correa, Hugo García-Compeán, Yulier Jiménez-Santana, Aldo Martínez-Merino, Leopoldo Pando Zayas and Roberto Santos-Silva for collaboration, discussions and the development of the ideas summarized here. I would like to thank useful discussions and comments from: Alejandro Cabo Bizet, Albrecht Klemm, Stefan Groot-Nibbelink, Martin Ro\^cek, Eric Sharpe, Hans Jockers, Kentaro Hori and David Tong. I would like to thank the Isaac Newton Institute for Mathematical Sciences, Cambridge, for support and hospitality during the programme ``Black holes: bridges between number theory and holographic quantum information" where work on this paper was undertaken. This work was supported by EPSRC grant no EP/R014604/1.
I thank the support of the University of Guanajuato grants CIIC 251/2024 
"Teorías efectivas de cuerdas y exploraciones de aprendizaje de máquina" and CIIC
224/2023 "Conjeturas de gravedad cuántica y el paisaje de la teoría de cuerdas",  CONAHCyT Grant  A-1-S-37752 “Teorías efectivas de cuerdas y sus aplicaciones
a la física de partículas y cosmología”. I would like to acknowledge 
support from the ICTP through the Associates Programme (2023-2028).
I would also like to thank the Simons Center for Geometry and Physics and the organizers and participants of the conference ``Gauged Linear Sigma Models @ 30" where this work was presented and discussed.





\end{document}